\documentclass{iau}

\usepackage{amsmath}
\usepackage{graphicx}
\usepackage{multirow}

\begin{document}

\lefttitle{Roth, Jacoby, Ciardullo, Soemitro \& Arnaboldi}
\righttitle{Integral Field Spectroscopy: a disruptive innovation for the PNLF}

\jnlDoiYr{2023}
\doival{10.1017/xxxxx}
\volno{384}

\aopheadtitle{Proceedings IAU Symposium}
\editors{O. De Marco, A. Zijlstra, R. Szczerba, eds.}

\title{Integral Field Spectroscopy: a disruptive innovation for observations of Planetary Nebulae and the PNLF}

\author{Martin Roth$^1$, George H. Jacoby$^2$, Robin Ciardullo$^3$, Azlizan Soemitro$^1$, \\Peter M. Weilbacher$^1$, Magda Arnaboldi$^4$}
\affiliation{$^1$Leibniz-Institut f\"ur Astrophysik Potsdam (AIP), Germany}
\affiliation{$^2$NSF's NOIRLab, U.S.A.}
\affiliation{$^3$The Penn State University, U.S.A.}
\affiliation{$^4$European Southern Observatory}

\begin{abstract}
A quarter of a century has passed since the observing technique of integral field spectroscopy (IFS) was first applied to planetary nebulae (PNe). Progress after the early experiments was relatively slow, mainly because of the limited field-of-view (FoV) of first generation instruments.  With the advent of MUSE at the ESO Very Large Telescope, this situation has changed. MUSE is a wide field-of-view, high angular resolution, one-octave spanning optical integral field spectrograph with high throughput. Its major science mission has enabled an unprecedented sensitive search for Ly$\alpha$ emitting galaxies at redshift up to z=6.5. This unique property can be utilized for faint objects at low redshift as well. It has been demonstrated that MUSE is an ideal instrument to detect and measure extragalactic PNe with high photometric accuracy down to very faint magnitudes out to distances of 30 Mpc, even within high surface brightness regions of their host galaxies. When coupled with a differential emission line filtering (DELF) technique, MUSE becomes far superior to conventional narrow-band imaging, and therefore MUSE is ideal for accurate Planetary Nebula Luminosity Function (PNLF) distance determinations.  MUSE enables the PNLF to become a competitive tool for an independent measure of the Hubble constant, and stellar population studies of the host galaxies that present a sufficiently large number of PNe.
\end{abstract}


\maketitle

\section{Introduction}
\label{Intro}
This paper describes the breakthrough for the spectrophotometry of Planetary Nebulae, specifically extragalactic PNe, that has been enabled with the technique of integral field spectroscopy. Most notably, the performance of MUSE, the Multi Unit Spectral Explorer, at the ESO Very Large Telescope (VLT) will be explained, as will how spectral imaging with IFS has essentially superseded narrow-band imaging, the classical technique for the determination of the Planetary Nebula Luminosity Function (PNLF), for distance measurements beyond  $\sim$20~Mpc. Not by coincidence, IAU Symposium 384 has yielded several more presentations concerned with the subject, e.g. \cite{2023IAUS384Jacoby}, \cite{2023IAUS384Soemitro}, \cite{2023IAUS384Bhattacharya}, \cite{2023IAUS384Hartke}. MUSE has also shown to be a powerful tool for spatially resolved plasma diagnostics of Galactic PNe, e.g., \cite{2023IAUS384Montero}. However this subject deserves a full account on its own merits and is beyond the scope of this review.

We will begin with a somewhat selective, brief history of IFS, an introduction to the MUSE instrument, a description of MUSE data cubes, and a discussion of the data analysis. We will then present a proof of concept and benchmark tests, and conclude with a summary and an outlook to the intended future development of the technique.

\begin{figure}[th!]
  \centerline{\includegraphics[scale=.13]{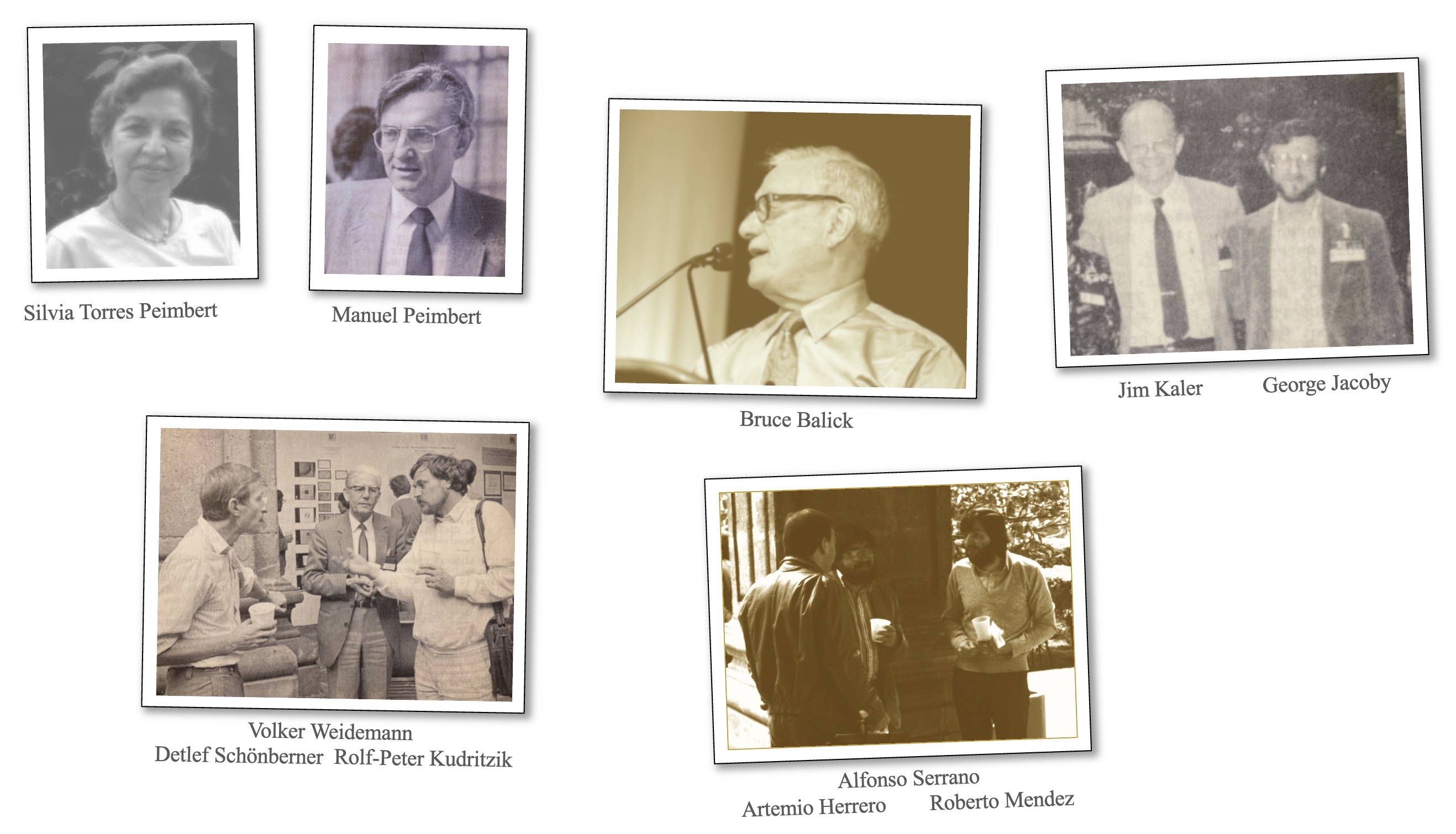}}
  \caption{A reminiscence of IAU Symposium 131 from 1987 in Mexico City, organized by Silvia Torres-Peimbert and Manuel Peimbert, that publicly marked the birth of the PNLF as a standard candle.}
  \label{Mexico}
\end{figure}

\pagebreak 

Before starting the discussion, it is useful and appropriate to put the problem into historical context and recall some key issues that matter for precision spectrophotometry, which is a prerequisite for accurate distance determinations with the PNLF. It is hoped that the reader will tolerate a slight personal twist when pointing out that a previous IAU Symposium on Planetary Nebulae, No. 131, that was organized on October 5-9, 1987 in Mexico City by Silvia Torres Peimbert, Manuel Peimbert, and Jim Kaler (SOC chair), can be understood as a landmark. It occurred when CCD detectors were beginning to enable precision spectrophotometry, and the less efficient technologies of photomultipliers and image tubes were being retired. In fact, the presentations by Holland Ford \citep{1989IAUS..131..335F} and George Jacoby \citep{1989IAUS..131..357J}  signaled the launch of the subsequent success story of the PNLF and the large number of studies that have extended over more than three decades since then -- essentially until most of the galaxies within reach of narrow-band imaging on 4-8m class telescopes were completed \citep{2022FrASS...9.6326C}. IAU 131 was also the time where CCD cameras revolutionized the study of PN morphology through narrow-band imaging as participants will recall from a memorable talk by Bruce Balick \citep{1989IAUS..131...83B} .

The observational problem can be understood as follows. Until the change of paradigm in the late 1980s, measuring emission line fluxes had conventionally been accomplished either with photoelectric aperture photometers, equipped with narrow-band interference filters, or spectrally dispersed spectrographs/scanners. The former were eventually replaced by narrow-band imaging CCD cameras \citep{1987PASP...99..672J} , but the latter are still in use with modern instrumentation. What was probably less rigorously absorbed by the community is the understanding how the accuracy of emission line flux measurements depends on observing conditions and details of the flux calibration. The fact that flux standards are typically hot stars, preferentially with featureless spectral energy distributions, introduces a complication related to the point spread function (PSF) of unresolved point sources: any aperture presented by the apparatus measuring stellar flux, whether it is the diaphragm of a photoelectric photometer, or the slit of a spectrograph, inevitably truncates the extended wings of the PSF, such that only a fraction of the light  coming from the star will be registered on the detector. Selecting a very large aperture diameter, or opening the slit jaws may minimize the light loss, but at the expense of the inclusion of background light, or reduced spectral resolution, respectively. Generally, one must ensure that variable atmospheric seeing does not result in the object and flux standard observations having different PSF widths that would result in systematic calibration errors. The paper introducing two-dimensional spectrophotometry with the panoramic CCD detector by \cite{1987PASP...99..672J} explains how the effects of a fixed aperture can be circumvented with the calibration techniques of CCD photometry. This pioneering work can be understood as setting the stage for what was exploited fully in subsequent PNLF studies. 

An important paper that has pioneered spectrophotometry of extragalactic PNe was presented by \cite{1993ApJ...417..209J}, who investigated in detail the various aperture effects, such as slit losses, seeing effects, differential atmospheric dispersion, etc. It is unfortunate that this presentation of fundamental technical concerns has not received the attention it deserves, but it is highly recommended for study. For illustration, see \cite{2023A&A...671A.142S} who discuss an example in the literature where ignoring those principles has likely led to a systematic bias in the PNLF photometry for the nearby galaxy NGC\,300.

As will be seen later when we introduce the observing technique of IFS, concerns related to aperture effects vanish as they can be treated by analysis software in the picture of a data cube.

\section{A brief history of integral field spectroscopy}
\label{History}

As an anecdotal subtlety, it is worthwhile to cite from the conclusion section of \cite{1987PASP...99..672J} who proposed an extension of their scheme of narrow-band filters for CCD photometry of PNe, and with foresight contemplated: "{\em Ideally, one would like a spectrograph having thousands of input apertures which can be configured to a packed array of serveral hundred per side to sample the sky at arc-sec intervals. Some fiber-optic systems now in use are testing this technique \citep{1986BAAS...18..951B}.}"

\begin{figure}[th!]
  \centerline{\includegraphics[scale=.43]{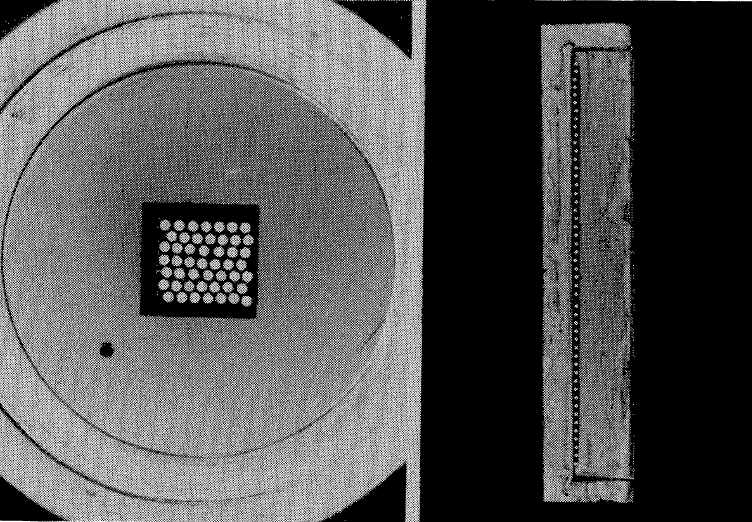}}
  \caption{The pioneering fiber bundle integral field unit DensePak II for the KPNO 4m Telescope. The unit featured 49 fibers and subtended $16 \times19$~arcsec$^2$ on the sky (from \cite{1988ASPC....3..113B}.}
  \label{DensePak}
\end{figure}

The statement was inspired by early experiments with panoramic IFS by French groups \citep{1987AnPh...12..207V,1988ESOC...30.1185B}, and at the Kitt Peak National Observatory 4m Telescope \citep{1988ASPC....3..113B}. Fig.~\ref{DensePak} shows an example of a relatively simple fiber bundle integral field unit (IFU), whose purpose was to resample the focal plane (left) onto a fiber pseudo-slit (right), that was fed to a long-slit spectrograph. While the former groups were primarily interested in the study of extragalactic problems, e.g., stellar kinematics in the nuclei of galaxies, or gravitationally lensed QSOs, it was another variant of the DensePak IFU with a total of 91 fibers, attached to the WIYN 3.5 m telescope and the fiber-fed bench spectrograph HYDRA in Echelle configuration, that is arguably the first IFS that was ever used for the study of PNe: 
\cite{1998AJ....116.1367J} investigated the spatially resolved expansion velocity field and the expansion age of the born-gain PN V4334 Sgr (Sakurai's object). Other works on Galactic PNe based on IFS include, e.g., \cite{2005ApJ...628L.139M} and \cite{2008AA...486..545S}, both of which studied the faint surface brightness haloes of PNe, or \cite{2010A&A...512A..18S} with an investigation of an extremely metal poor galactic halo PN.

\begin{figure}[h!]
  \centerline{\includegraphics[scale=.142]{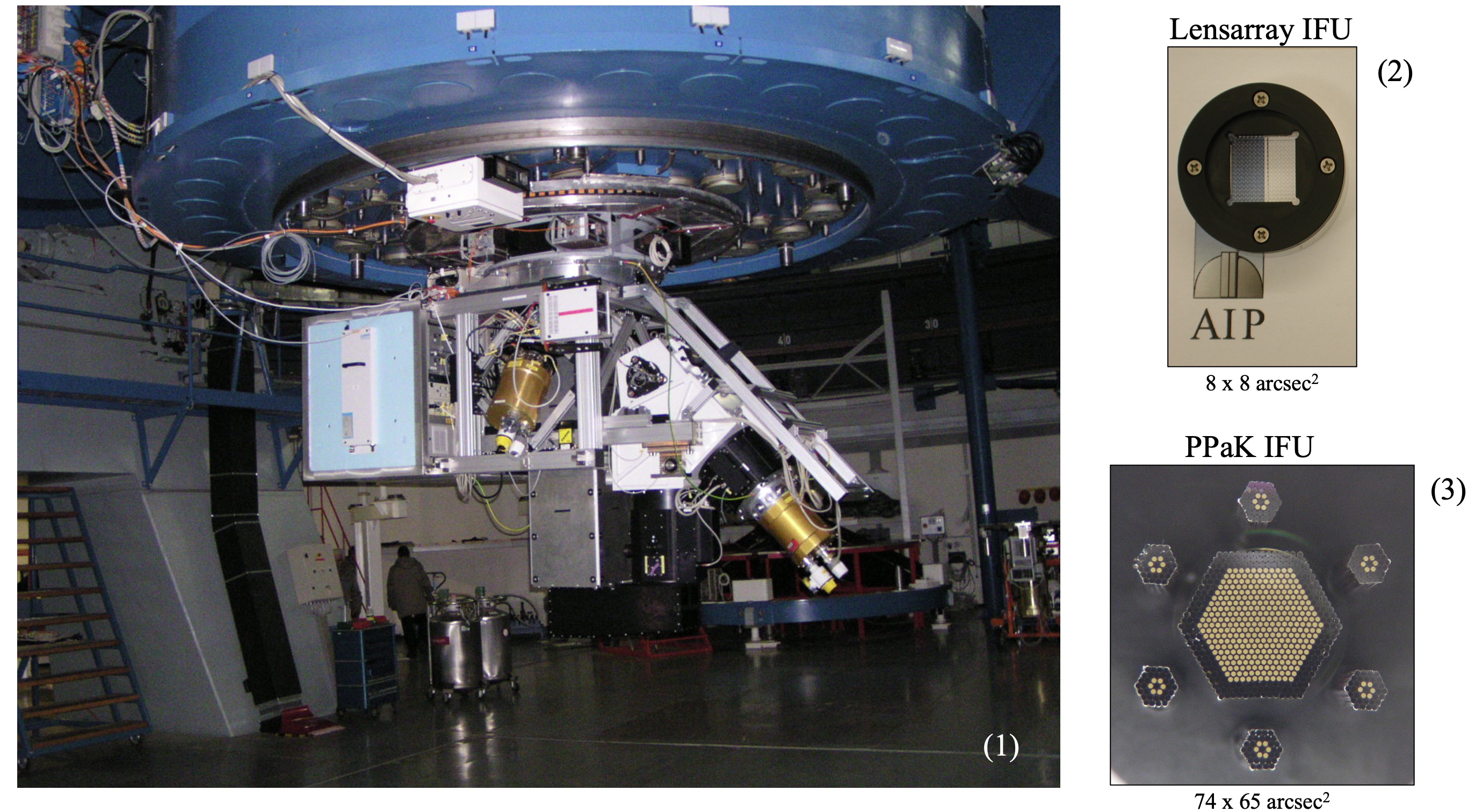}}
  \caption{Left: PMAS instrument at the Cassegrain focus of the Calar Alto 3.5m Telescope (1). Right: square lens array IFU (2) with a standard FoV of $8\times8$~arcsec$^2$; fiber bundle IFU PPaK with hexagonal footprint and FoV of $74\times65$~arcsec$^2$, surrounded by 6 bundles to sample the sky background (3).}
  \label{PMAS}
\end{figure}

Coming back to the importance of aperture effects as discussed in Section~\ref{Intro}, it is useful to focus on a  first generation effort with IFS that was rigorously designed to enable precision spectrophotometry, as can be appreciated from the acronym PMAS: Potsdam Multi-Aperture Spectrophotometer \citep{1998SPIE.3355..798R}. Unlike some other  spectrographs that were retrofitted with an IFU, PMAS was specifically built and optimized for that purpose. It is a fully self-contained instrument, shown in Fig.~\ref{PMAS}, at the Cassegrain focus of the Calar Alto Observatory (CAHA) 3.5m telescope \citep{2005PASP..117..620R}. As a unique feature, it offers two IFU modes that are complementary in terms of angular resolution and FoV. The standard lens array IFU covers a FoV between $8\times8$~arcsec$^2$ and $16\times16$~arcsec$^2$ with a choice of three different magnifications, and Nyquist sampling to accomodate seeing of $1\ldots2$~arcsec FWHM; square lenses of projected 0.5~arcsec pitch are used for the highest angular resolution. The PPaK IFU is a bare fiber bundle that was retrofitted to the initial configuration in an off-axis position of the Cassegrain focal plane for the purpose of allowing an approximately 1~arcmin wide FoV with large spatial sampling elements (spaxels). The resulting $\approx 3$~arcsec fiber diameter on the sky was built to enable the observation of low surface brightness objects \citep{2006PASP..118..129K}. Another unique feature of PMAS is its high throughput and sensitivity down to the atmospheric cutoff in the ultraviolet, as demonstrated in a science verification run with observations of the high excitation PN NGC\,7027 \citep{2004AN....325..147R}.

Analogous to the slicer-based IFS instrument E3D \citep{1996AAS..119..531W}, PMAS was initially intended to be a traveling experiment and designed with the goal of being used as a visitor instrument on different telescopes. However, after commissioning at the CAHA 3.5m telescope in 2001, the community interest was found to be so high that it has been there as a shared risk common user instrument ever since. PMAS has been used for a diversity of problems, e.g.,  Milky Way Herbig-Haro objects \citep{2008MNRAS.384..464L},
HII regions in nearby galaxies \citep{2013MNRAS.430..472L},
supernovae \citep{2003AA...401..479C},
ultra-luminous X-ray sources \citep{2005AA...431..847L},
extremely metal-poor starburst galaxies \citep{2016MNRAS.459.2992K},
blue compact dwarf galaxies \citep{2010AA...520A..90C},
luminous infrared galaxies \citep{2009AA...506.1541A},
multiply lensed QSOs \citep{2003AA...408..455W},
DLA galaxies and their foreground absorbers \citep{2004AA...417..487C},
GRB host galaxies \citep{2009AA...497..729F},
and surveys like DiskMass \citep{2010ApJ...716..198B} and CALIFA \citep{2012AA...538A...8S},
and many more.

\begin{figure}[t!]
   \centerline{\hspace{3mm}  \includegraphics[scale=.15]{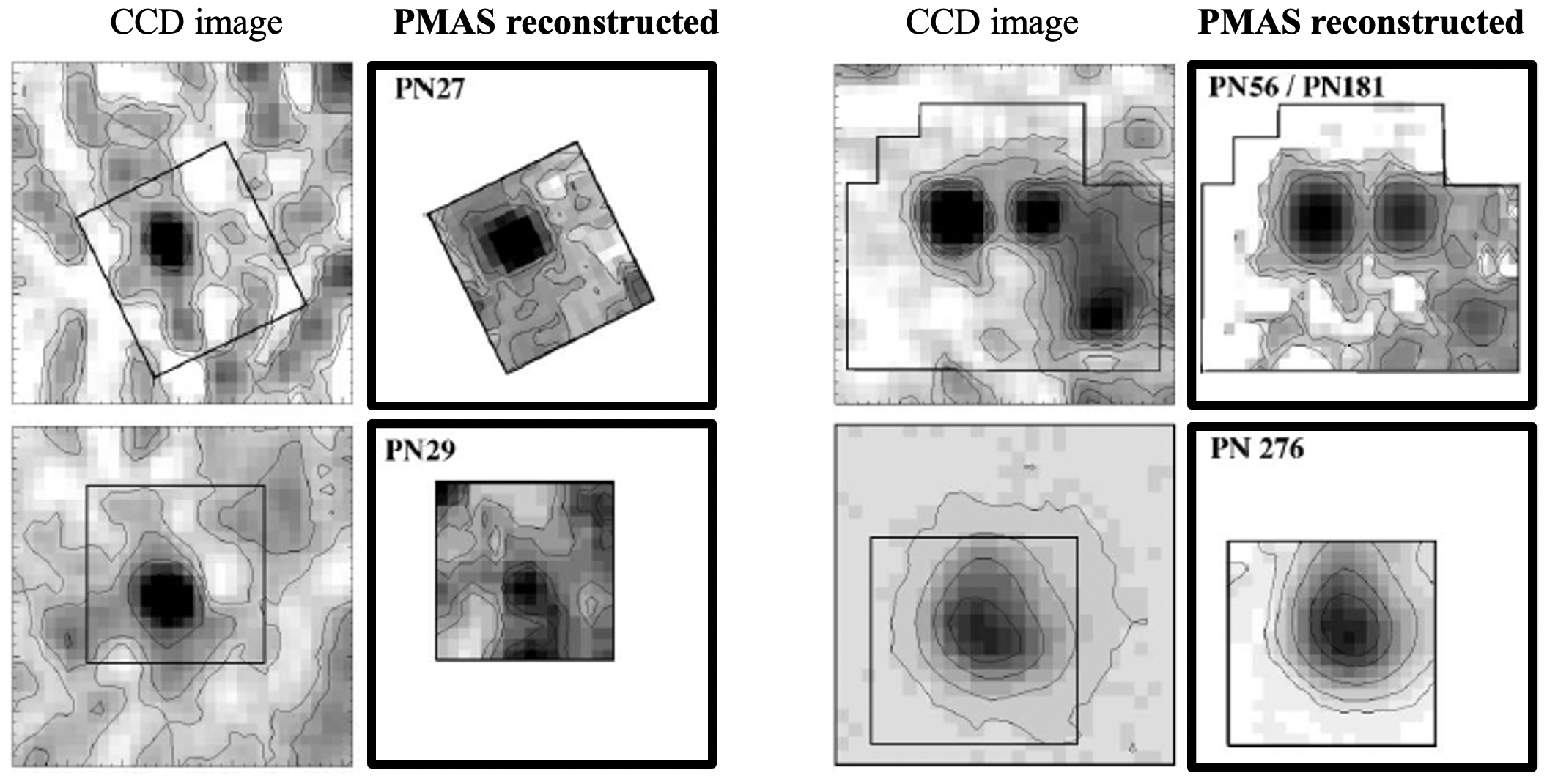}}
   \vspace{5mm}
   \centerline{\includegraphics[scale=.15]{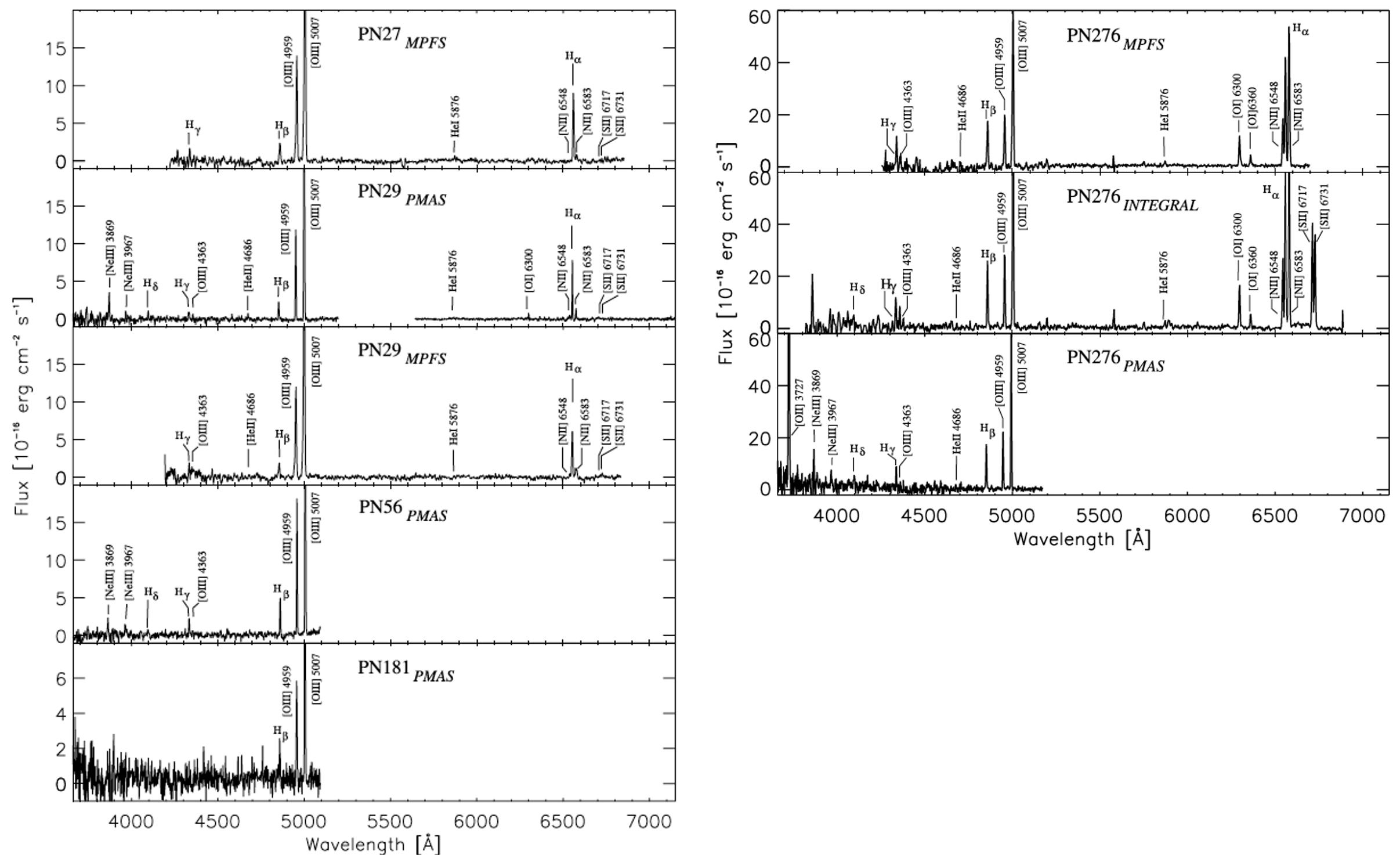}}
   \caption{PNe in M31 observed with IFS. Adapted from \cite{2004ApJ...603..531R}.
   Top: the bold outlined frames show  reconstructed maps obtained from PMAS data cubes in [O\,III] 
    $\lambda5007$\AA ~for PN27, PN29, PN56, PN181, and PN276 from the sample of \cite{1989ApJ...339...53C};
    Fabry Perot CCD images are shown on the left for comparison (see explanation in the text). Bottom: corresponding    
    spectra obtained with PMAS, MPFS at the Selentchuk 6m telescope, and with INTEGRAL at the WHT. PN276 is  
    clearly identifiable as a supernova remnant.
   }
   \label{PMAS-PN}
\end{figure}

The seeing-limited spatial sampling of PMAS has been utilized to demonstrate crowded field  IFS for stars in globular clusters  and the validation of the PampelMuse code, that can be understood as an analogue to DAOPHOT \citep{1987PASP...99..191S} for data cubes \citep{2013AA...549A..71K}. This latter technique has become an ''industry'', used on objects ranging from Milky Way globular clusters out to local volume galaxies \citep{2019AN....340..989R}, and has been applied to data from MUSE at the ESO VLT (see below). This is a satisfying confirmation of the PMAS science case, presented in \cite{1998ASPC..152..168R}. Two other science cases for PMAS included the study of spatially extended Milky Way PNe, and extragalactic PNe, that are, akin to stars, observed as point sources. The idea was driven by the argument that in the absence of aperture effects, IFS was thought to be an ideal tool for spectrophotometry.

Despite the small number of spaxels (256) and the quite limited FoV, a proof of concept was undertaken with PNe in the bulge of the Andromeda galaxy, selected from the sample in \cite{1989ApJ...339...53C}, who had obtained narrow-band imaging photometry in [O\,III]  $\lambda5007$\AA\ and established the PNLF for M31. The results from observations with three integral field spectrographs were published in \cite{2004ApJ...603..531R}: these included early observations with the MPFS instrument in the prime focus of the Selentchuk 6m telescope \citep{2000AJ....119..126A}, from the INTEGRAL IFU \citep{1998ASPC..152..149A}, coupled to the WYFFOS spectrograph at the William Herschel Telescope (WHT), and from the first PMAS science observations that were obtained after commissioning and science verification. Fig.~\ref{PMAS-PN} shows example maps and spectra that were extracted from the data cubes that had been obtained from these observations. The validity of the imaging capabilities was independently checked with narrow-band images, obtained with the Fabry Perot camera in the prime focus of the CAHA 3.5m telescope. 

Incidentally, it was discovered that one of the objects, PN276, is not a point source as it exhibits a roughly triangular geometry, and displays strong low excitation lines of [O\;I], [S\;II], and [N\;II]. These lines are characteristic for supernova remnants (SNR). 

Moreover, \cite{2004ApJ...603..531R} were able to demonstrate that a PSF-fitting technique, which is common-place in crowded field CCD photometry, is useful for distinguishing intrinsic PN emission line strengths from contamination due to diffuse gas emission in the environment of the point source. More observations in Local Group galaxies have shown the usefulness of this approach, albeit with a low efficiency, since the initial PMAS configuration was essentially using only half of the spectrograph's spectral range (the instrument's 2K$\times$4K CCD was later replaced by a 4K$\times$4K chip \citep{2010SPIE.7742E..09R}). Nonetheless, another misidentification of a PN candidate was reported in \citep{2006pnbm.conf..239R}. Still, beyond the proof of principle, a more productive use of IFS for the observation of extragalactic PNe had to await the arrival of a large panoramic integral field spectrograph with high angular resolution - such as MUSE.

\section{MUSE, the Multi Unit Spectroscopic Explorer}
\label{MUSE}

MUSE is a 2nd generation VLT instrument that was built between 2004 and commissioning in 2014 \citep{2010SPIE.7735E..08B}, that is mounted on the Nasmyth platform of Unit Telescope~4 (UT4) of the Very Large Telescope (VLT) of the European Southern Observatory (Fig.~\ref{MUSE}). As a survey instrument built to address faint high redshift galaxies in the optical wavelength domain, MUSE is entirely devoted to single mode IFS. The instrument achieves the best possible stability by avoiding many different observing modes and the usually required moving parts. The primary mirror of UT4 was the last one manufactured for the VLT and exhibits the the smallest wavefront errors due to imperfections of the polishing procedure. It is therefore the telescope of choice for implementation of a ground layer adaptive optics system, called GALACSI \citep{2006NewAR..49..618S}. GALACSI is available for MUSE, both in the standard wide field mode, covering a FoV of $1\times1$~arcmin$^2$ with square spaxels of $0.2\times0.2$~arcsec$^2$, and the narrow field mode with a FoV of $7.5\times7.5$~arcsec$^2$ and $0.025\times0.025$~arcsec$^2$ spaxels. The wavelength range is 465 - 930~nm at a spectral resolution of R$=2000\ldots4000$. 

As opposed to the fiber-based IFUs shown in Fig.~\ref{DensePak} and Fig.~\ref{PMAS}, MUSE achieves its 90,000 spaxels (and spectra) through a modular design, involving 24 identical spectrographs, each of which is equipped with an image slicer to ensure the densest possible packing of spectra on the detector. The 24 spectrographs are fed from the focal plane of the telescope through a derotator and a field splitter fore optics such that a free space optical system from the telescope pupil to the detector is realized. Each detector pixel is mapped onto a particular spot on the sky and a particular wavelength bin. The data reduction pipeline collects this mapping in a so-called pixel table from which, after appropriate astrometric and wavelength calibration, a final data cube is formed in one single rebinning step \citep {2020A&A...641A..28W}. 

\begin{figure}[t!]
   \centerline{\includegraphics[scale=.135]{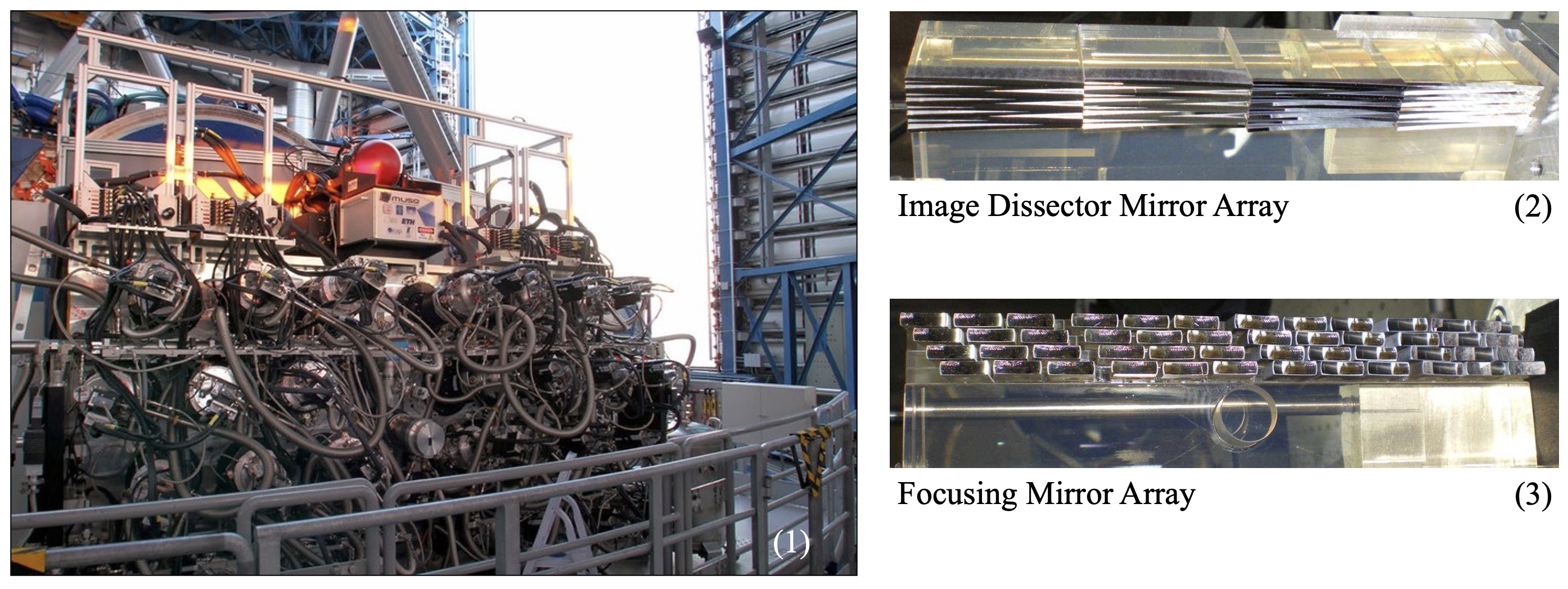}}
   \caption{MUSE at the ESO VLT. The complex arrangement of 24 spectrographs on the Nasmyth platform of UT4 is shown in (1), credit: Ghaouti Hansali (CRAL). The 24 spectrographs are fed by image slicers whose purpose is to map the input sky field into a staggered set of mini-slits. Each image slicer consists of an arrangement of an image dissector array (2) with a focusing mirror array (3), that sends an ordered set of sliced sky area to its associated spectrograph module. Credit: Florence Laurent, CRAL.}
   \label{MUSE}
\end{figure}

The slicer concept is depicted in Fig.~\ref{MUSE}, (2) and (3). It consists of an image dissector array (IDA), made out of 4 identical stacks of 12 slices, which cut the input FoV into thin, narrow strips, thus redirecting the beams in different directions. The telescope pupil is reimaged onto different intermediate pupils that are arranged in a staggered geometry, allowing for straylight removal through pupil stops. The light is further directed to a matching focusing mirror array (FMA), that finally sends the light into the spectrograph collimator for dispersion through a volume phase holographic grating and subsequent imaging onto the CCD detector. A full description of this critical device and its performance is found in \cite{2008SPIE.7018E..0JL}.

The complex design of MUSE and its performance, including stringent priorities to optimize throughput, set this instrument aside from other competing large integral field spectrographs on 8-10m class telescopes, like KCWI at Keck \citep{2018ApJ...864...93M}, or VIRUS at the Hobby Eberly Telescope \citep {2021AJ....162..298H}. The large FoV, high image quality with ground layer adaptive optics support, and high throughput make this instrument unique, to the effect that it has become the most demanded and productive facility of the VLT (e.g., 170 refereed papers in 2022)\footnote[1]{https://www.eso.org/sci/php/libraries/pubstats/}.

The science objectives that were identified during MUSE's Phase A study include, first and foremost, the study of the formation of galaxies at high redshift, e.g., Ly$\alpha$ emitters, fluorescent emission and the cosmic web, reionisation, feedback processes and galaxy formation, ultra-deep surveys using strong gravitational lensing, resolved spectroscopy at intermediate redshifts. Also listed are the investigation of nearby galaxies (e.g., kinematics and stellar populations, interacting galaxies, star formation in nearby galaxies, the study of stars and resolved stellar populations, early stages of stellar evolution), and crowded field spectroscopy in the Milky Way, Local Group galaxies, and beyond. 

It is beyond the scope or this paper to review the wealth of scientific outcome obtained with MUSE. However for the subject of spectrophotometry of faint extragalactic PNe, it is useful to highlight the sensitivity obtained with a very deep exposure of 140 h, obtained in the Hubble Ultra-Deep Field, aka the MUSE Extremely Deep Field 
\citep[MXDF;][]{2021A&A...647A.107B}. At a wavelength of 7000~\AA, this exposure has enabled an unprecedented $5\sigma$ surface brightness detection limit of $1.3\times10^{-19}$ erg cm$^{-2}$s$^{-1}$ arcsec$^{-2}$ for diffuse extended Ly$\alpha$ emission, while the $5\sigma$ point source limiting flux is  $2.3\times10^{-19}$ erg cm$^{-2}$s$^{-1}$ for the same wavelength. Thanks to the adaptive optics system, the image quality of the combined dataset was excellent, with a FWHM of 0.6~arcsec at 4700~\AA, and 0.4~arcsec at 9300~\AA. 

The ability to accumulate long exposures to achieve this level of sensitivity is due to the instrument's stability that was  a design goal from the outset. It can of course be utilized for faint objects at low redshift as well, which was indeed envisaged by a proposed study in the nearby galaxy NGC\,300 as part of the Phase A science case mentioned above. The MUSE collaboration has therefore conducted observations in selected fields of NGC\,300 as part of the guaranteed observing time program with run IDs 094.D-0116(A), 094.D-0116(B), 095.D-0173(A), 097.D-0348(A), 0102.B-0317(A) between September 2014 and October 2018, entitled "{\em A study of the faint end of the planetary nebulae luminosity function of NGC 300}", PI Roth. The program produced data for a total of 9 fields with total exposure times of 1.5~h each (under photometric conditions), and mostly with excellent image quality (ground layer adaptive optics support during the 2018 run). 

A pilot study of the initial 7 pointings was published by \cite{2018A&A...618A...3R}  and presented results from the technique of crowded field IFS as described in Section~\ref{History}. This paper presented a catalog of luminous stars, rare objects such as Wolf-Rayet stars, Be stars, carbon stars, symbiotic star candidates, PNe, H II regions, SNR, giant shells, peculiar diffuse and filamentary emission line objects, and background galaxies, along with their spectra.

\begin{figure}[h!]
   \centerline{\includegraphics[scale=.12]{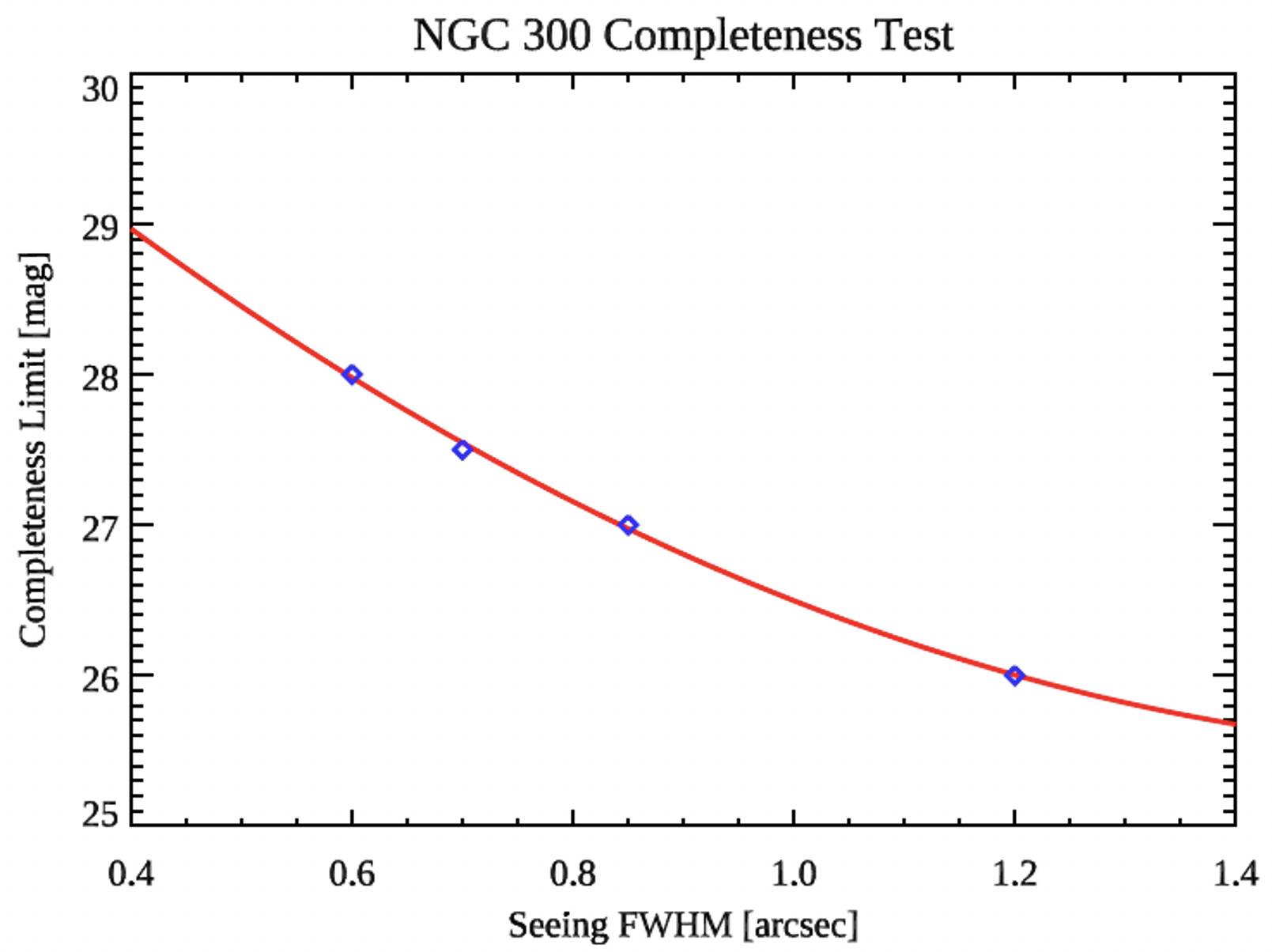}}
   \caption{Completeness of PN detections at 95\% in MUSE data cubes observed in NGC\,300 as a function of seeing 
   FWHM.  The total exposure time for each observation was 1.5~h. From \cite{2018A&A...618A...3R}.}
   \label{PN_completeness}
\end{figure}

By mimicking PNe with artificial emission line point sources that were embedded in the observed data cubes, it was demonstrated that 95\% completeness could be achieved at a limiting magnitude of m5007 = 28~mag when the image quality was excellent (0.6~arcsec FWHM), and m5007 = 26~mag when the seeing was mediocre (1.2~arcsec FWHM). The definition of monochromatic [OIII] magnitudes m5007 is given in \cite{1989ApJ...339...53C}, and is roughly equal to the magnitude of the object through a broadband $V$ filter. These very encouraging results have prompted the further development of MUSE spectrophotometry of extragalactic PNe as described below. 

To date, the rich dataset from NGC\,300 has spawned three more papers: \cite{2022A&A...658A.117G}  on the quantitative spectroscopy of BA-type supergiants, \cite{2022A&A...668A..74M} on extremely faint HII regions and diffuse ionized gas, and \cite{2023A&A...671A.142S} on the PNLF. The latter work is also discussed in these proceedings \citep{2023IAUS384Soemitro}. The improved analysis of the partially resolved stellar population is ongoing, and more papers are in preparation.

\section{Differential Emission Line Filter (DELF)}
\label{DELF}

\begin{figure}[h!]
   \centerline{\includegraphics[scale=.135]{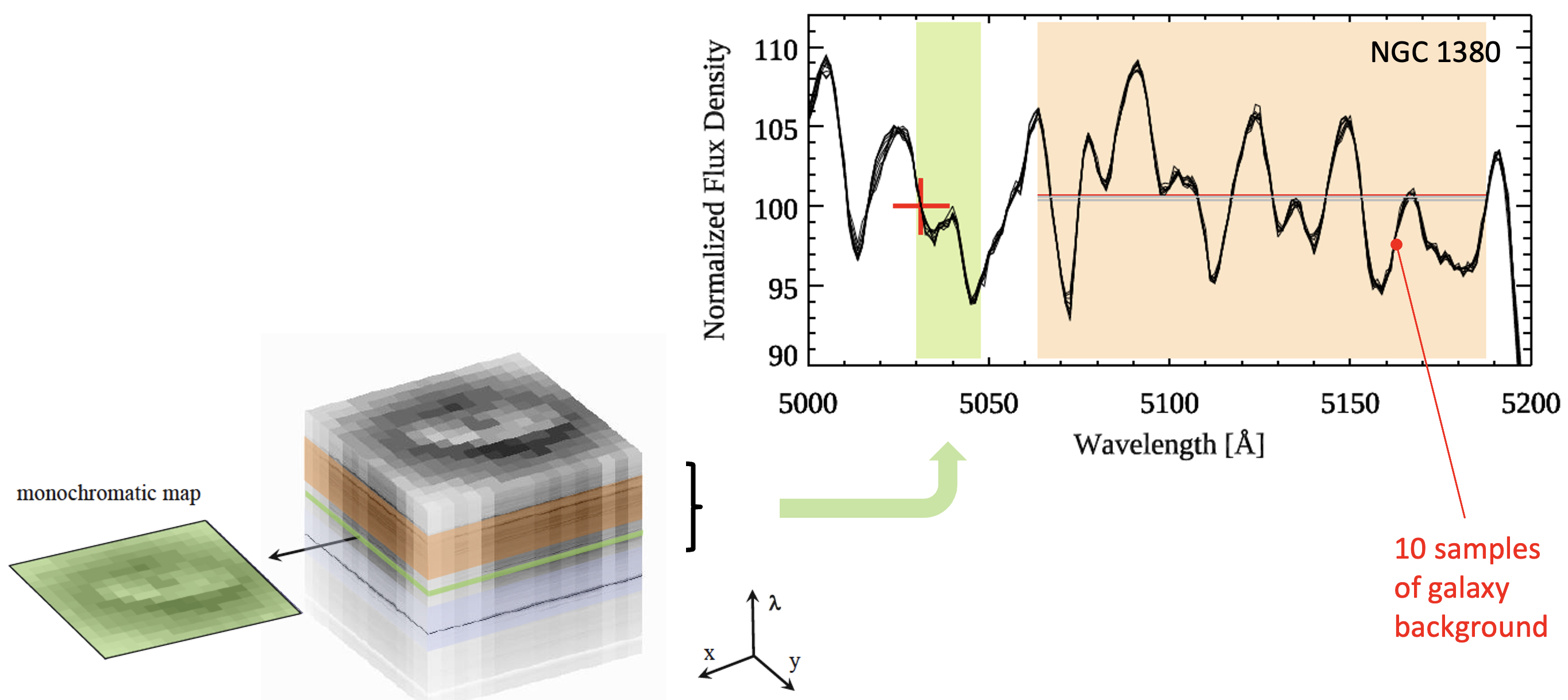}}
   \caption{The Differential Emission Line Filter for data cubes (DELF). Figure adapted from \cite{2021ApJ...916...21R}. For explanation, see text.}
   \label{DELF_scheme}
\end{figure}

As reported by \cite{2018A&A...618A...3R},  the extremely narrow 1.25~\AA\ bandwidth of a single MUSE data cube layer results in very high contrast for emission line objects, no matter whether the objects are point sources or extended surface brightness distributions. In their analysis, the residual continuum within the chosen $5\ldots6.25$~\AA\ wavelength intervals for the synthetic filters of important emission lines was subtracted with suitably chosen and scaled continuum nearby passbands. Fig.~\ref{DELF_scheme} illustrates how this scheme has been further developed to form a Differential Emission Line Filter (DELF) for the [OIII] $\lambda$5007~\AA\ line in MUSE data cubes.

The green and orange hues of the sketch are intended to illustrate passbands centered on the [OIII] emission line (on-band), and with a much wider bandwidth on the nearby continuum (off-band). For classical PNLF observations with narrow-band filters, the on-band and off-band exposures would be obtained one after another at different times during an observing night. 

The key point here is that a MUSE data cube provides on-band and off-band data obtained {\em simultaneously}, and under {\em exactly the same observing conditions}. The spectra shown in Fig.~\ref{DELF_scheme} depict an example obtained with deep observations of the lenticular galaxy NGC\,1380, featuring various metal absorption lines, and a peak-to-peak variation across the average of roughly 5\%. It must be noted that the plot contains the superposition of 10 spectra that were taken from various locations over the face of the galaxy. They are almost indistinguishable, and the variation from spectrum to spectrum is tiny. Also the averaged flux density of the continuum over the entire off-band shows a remarkably small variation on the order of 0.1\%.

We emphasize that the continuum background for emission line objects within the green on-band interval is affected by a blend of absorption lines that happen to show a variation of 5\% from the blue end to the red end, corresponding to a radial velocity interval of $\pm$250~km~s$^{-1}$, which is typical for a large galaxy. The central [OIII] wavelength of the most blue-shifted PN in this picture would correspond to the red cross. However, the {\em ratio between the intensity at a given wavelength} within the on-band and the {\em continuum average} of the off-band does not change from spectrum to spectrum. In other words: for each wavelength bin within the green passband, we can apply a calibrated subtraction of the average off-band, which is very well defined because of the large interval, and the small spectral variation. As a result, an accurate, internally calibrated continuum background subtraction is obtained on a spaxel-by-spaxel basis. In essence, this is the reason for the excellent performance of DELF.

A mathematical derivation for the performance of a DELF is presented in \cite{2021ApJ...916...21R}. In a sense, the DELF technique bears some similarities with the problem of accurate sky background subtraction for the detection of the faintest  high redshift galaxies before the advent of the Hubble Space Telescope. The search for background-limited galaxies by \cite{1990ASPC...10..292T} was only accomplished by achieving an accuracy of $10^{-4}$ of the sky background through the so-called "shift-and-stare" technique. This methodology overcame pixel-to-pixel sensitivity variations in the flatfield calibration thanks to the linearity and stability of the CCD detector. The advantage of MUSE is that the spaxel-to-spaxel variations cancel out immediately due to the inherent differential principle within a data cube, and no offset pointings are required.

\section{MUSE PNLF}
\label{PNLF}

\begin{figure}[h!]
   \centerline{\includegraphics[scale=.145]{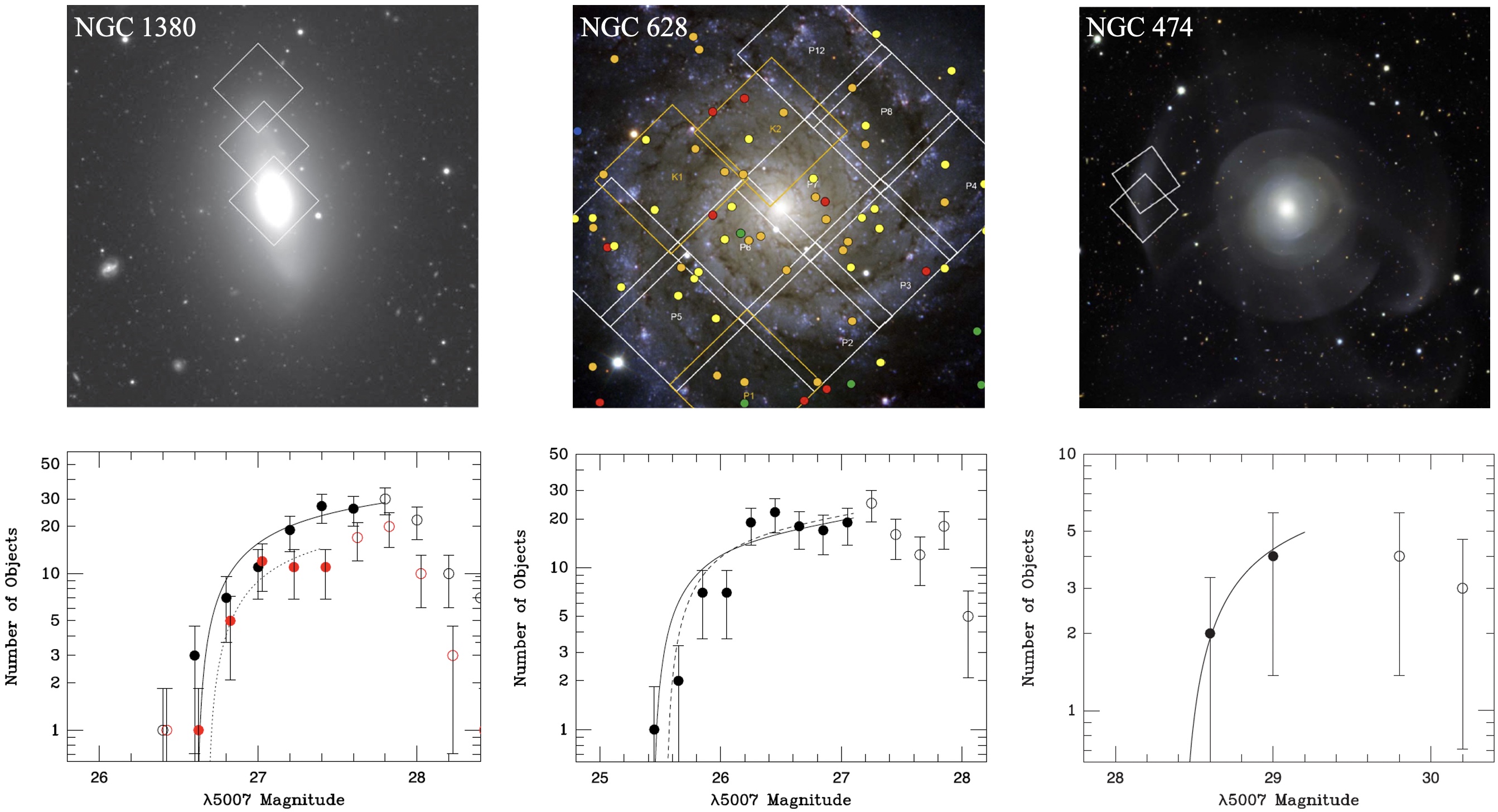}}
   \caption{Three benchmark galaxies used to validate the new approach of measuring the PNLF with MUSE and the DELF methodology. Adapted from \cite{2021ApJ...916...21R}.}
   \label{Benchmark}
\end{figure}

Given the favorable performance of the DELF technique, a proof of concept was undertaken using archival data to investigate whether MUSE spectrophotometry can improve the quality and range of PNLF distance determinations significantly beyond the state-of-the art with the classical narrow-band filter technique. The study, published in \cite{2021ApJ...916...21R}, discusses in detail the test of photometric precision, the calibration of aperture corrections, the additional benefits of discarding interlopers (SNR, HII regions, background galaxies) based on candidate spectra, the utility of avoiding systematic errors due to interference filter passband transmission changes, and the unique opportunity to discover and eliminate systematic errors caused by chance superpositions of PNe, see also \cite{2023ApJ...950...59C}. The results for three benchmark tests, namely NGC\,1380, NGC\,628, and NGC\,474, are illustrated in Fig.~\ref{Benchmark}. 

Comparison with other PNLF studies in the literature based on the same data but using other techniques demonstrates that DELF is superior in terms of signal-to-noise ratio and sensitivity. A more elaborate test with 20 galaxies of various Hubble types and distances has been published in \cite{2023arXiv230911603J} and is discussed in these proceedings as a promising independent technique to measure the Hubble constant \citep{2023IAUS384Jacoby}.

\section{Summary and Conclusions}
Integral field spectroscopy has been demonstrated to be the ideal observing technique for precision spectrophotometry: a prerequiste for accurate PNLF distance determinations. However, first generation instruments were offering too small a field-of-view to efficiently use them for that purpose. MUSE, with its wide field-of-view, excellent image quality, and high throughput at an 8m class telescope, has completely changed the situation. Thanks to the superior quality of MUSE spectrophometry, the DELF technique, and the associated flatfield accuracy, we are now able to accurately perform [OIII] $\lambda$5007~\AA\ photometry even in high surface brightness regions near the nuclei of galaxies, which contain a high surface density of planetary nebulae. PN surveys in these regions were previously impossible with conventional techniques. After a successful proof of concept study and the exploration of archival data for a sample of 20 galaxy test cases, it is now time for targeted observations and carefully planned calibrations that are optimized for MUSE PNLF determinations. The method holds great promise to provide a new, independent way of measuring the Hubble constant. Looking further into the future, the option of an IFU for WFOS at the TMT \citep{2022SPIE12184E..23S} could potentially increase the reach of PNLF distances out to $\sim$100~Mpc.


\begin{thebibliography}{}
\bibitem[Alonso-Herrero~{\it et al.}(2009)]{2009AA...506.1541A} Alonso-Herrero A., Garc{\'\i}a-Mar{\'\i}n M., Monreal-Ibero A., Colina L., Arribas S., Alfonso-Garz{\'o}n J., Labiano A., 2009, A\&A, 506, 1541. doi:10.1051/0004-6361/200911813
\bibitem[Arribas~{\it et al.}(1998)]{1998ASPC..152..149A} Arribas S., del Burgo C., Carter D., Cavaller L., Edwards R., Fuentes J., et al., 1998, ASPC, 152, 149
\bibitem[Bacon~{\it et al.}(1988)]{1988ESOC...30.1185B} Bacon R., Adam G., Baranne A., Court{\`e}s G., Dubet D., Dubois J.-P., et al., 1988, ESOC, 30, 1185
\bibitem[Bacon~{\it et al.}(2010)]{2010SPIE.7735E..08B} Bacon R., Accardo M., Adjali L., Anwand H., Bauer S., et al., 2010, SPIE, 7735, 773508. 
\bibitem[Bacon~{\it et al.}(2021)]{2021A&A...647A.107B} Bacon R., Mary D., Garel T., Blaizot J., Maseda M., Schaye J., Wisotzki L., et al., 2021, A\&A, 647, A107. doi:10.1051/0004-6361/202039887
\bibitem[Balick (1989)]{1989IAUS..131...83B} Balick B., 1989, IAUS, 131, 83
\bibitem[Bershady~{\it et al.}(2010)]{2010ApJ...716..198B}Bershady M.~A., Verheijen M.~A.~W., Swaters R.~A., Andersen D.~R., Westfall K.~B., Martinsson T., 2010, ApJ, 716, 198. doi:10.1088/0004-637X/716/1/198
\bibitem[Bhattacharya~{\it et al.}(2023)]{2023IAUS384Bhattacharya} Bhattacharya, S. et al., 2023, these proceedings
\bibitem[Barden \& Scott(1986)]{1986BAAS...18..951B} Barden S.~C., Scott K., 1986, BAAS
\bibitem[Barden \& Wade(1988)]{1988ASPC....3..113B} Barden S.~C., Wade R.~A., 1988, ASPC, 3, 113
\bibitem[Cairos~{\it et al.}(2010)]{2010AA...520A..90C} Cair{\'o}s L.~M., Caon N., Zurita C., Kehrig C., Roth M., Weilbacher P., 2010, A\&A, 520, A90. doi:10.1051/0004-6361/201014004
\bibitem[Chase~{\it et al.}(2023)]{2023ApJ...950...59C} Chase O., Ciardullo R., Roth M.~M., Jacoby G.~H., 2023, ApJ, 950, 59. doi:10.3847/1538-4357/acc9bd
\bibitem[Ciardullo~{\it et al.}(1989)]{1989ApJ...339...53C} Ciardullo R., Jacoby G.~H., Ford H.~C., Neill J.~D., 1989, ApJ, 339, 53. doi:10.1086/167275
\bibitem[Ciardullo (2022)]{2022FrASS...9.6326C}Ciardullo R., 2022, FrASS, 9, 896326
\bibitem[Christensen~{\it et al.}(2003)]{2003AA...401..479C} Christensen L., Becker T., Jahnke K., Kelz A., Roth M.~M., S{\'a}nchez S.~F., Wisotzki L., 2003, A\&A, 401, 479. doi:10.1051/0004-6361:20030015
\bibitem[Christensen~{\it et al.}(2004)]{2004AA...417..487C} Christensen L., S{\'a}nchez S.~F., Jahnke K., Becker T., Wisotzki L., Kelz A., Popovi{\'c} L. {\v{C}}., et al., 2004, A\&A, 417, 487. doi:10.1051/0004-6361:20034371
\bibitem[Ferrero~{\it et al.}(2009)]{2009AA...497..729F} Ferrero P., Klose S., Kann D.~A., Savaglio S., Schulze S., Palazzi E., Maiorano E., et al., 2009, A\&A, 497, 729. doi:10.1051/0004-6361/200809980
\bibitem[Ford~{\it et al.}(1989)]{1989IAUS..131..335F} Ford H.~C., Ciardullo R., Jacoby G.~H., Hui X., 1989, IAUS, 131, 335
\bibitem[Micheva~{\it et al.}(2022)]{2022A&A...668A..74M} Micheva G., Roth M.~M., Weilbacher P.~M., Morisset C., Castro N., Monreal Ibero A., Soemitro A.~A., et al., 2022, A\&A, 668, A74. doi:10.1051/0004-6361/202244017
\bibitem[Gonz{\'a}lez-Tor{\`a}~{\it et al.}(2022)]{2022A&A...658A.117G} Gonz{\'a}lez-Tor{\`a} G., Urbaneja M.~A., Przybilla N., Dreizler S., Roth M.~M., Kamann S., Castro N., 2022, A\&A, 658, A117. doi:10.1051/0004-6361/202142372
\bibitem[Hartke~{\it et al.}(2023)]{2023IAUS384Hartke} Hartke J et al., 2023, these proceedings
\bibitem[Hill~{\it et al.}(2021)]{2021AJ....162..298H} Hill G.~J., Lee H., MacQueen P.~J., Kelz A., Drory N., Vattiat B.~L., Good J.~M., et al., 2021, AJ, 162, 298. doi:10.3847/1538-3881/ac2c02
\bibitem[Jacoby~{\it et al.}(1987)]{1987PASP...99..672J} Jacoby G.~H., Quigley R.~J., Africano J.~L., 1987, PASP, 99, 672
\bibitem[Jacoby \& Kaler(1993)]{1993ApJ...417..209J} Jacoby G.~H., Kaler J.~B., 1993, ApJ, 417, 209
\bibitem[Jacoby~{\it et al.}(1998)]{1998AJ....116.1367J} Jacoby G.~H., De Marco O., Sawyer D.~G., 1998, AJ, 116, 1367
\bibitem[Jacoby (1989)]{1989IAUS..131..357J} Jacoby G., Ford H.~C., Ciardullo R., 1989, IAUS, 131, 357
\bibitem[Jacoby~{\it et al.}(2023a)]{2023arXiv230911603J} Jacoby G.~H., Ciardullo R., Roth M.~M., Arnaboldi M., Weilbacher P.~M., 2023, arXiv, arXiv:2309.11603. doi:10.48550/arXiv.2309.11603
\bibitem[Jacoby~{\it et al.}(2023b)]{2023IAUS384Jacoby} Jacoby G.~H. et al., 2023, these proceedings
\bibitem[Kamann~{\it et al.}(2013)]{2013AA...549A..71K} Kamann S., Wisotzki L., Roth M.~M., 2013, A\&A, 549, A71. doi:10.1051/0004-6361/201220476
\bibitem[Kehrig~{\it et al.}(2016)]{2016MNRAS.459.2992K} Kehrig C., V{\'\i}lchez J.~M., P{\'e}rez-Montero E., Iglesias-P{\'a}ramo J., Hern{\'a}ndez-Fern{\'a}ndez J.~D., Duarte Puertas S., Brinchmann J., et al., 2016, MNRAS, 459, 2992. doi:10.1093/mnras/stw806
\bibitem[Kelz~{\it et al.}(2006)]{2006PASP..118..129K} Kelz A., Verheijen M.~A.~W., Roth M.~M., Bauer S.~M., Becker T., Paschke J., Popow E., et al., 2006, PASP, 118, 129. doi:10.1086/497455
\bibitem[Laurent~{\it et al.}(2008)]{2008SPIE.7018E..0JL} Laurent F., Renault E., Kosmalski J., Adjali L., Boudon D., Bacon R., Caillier P., et al., 2008, SPIE, 7018, 70180J. doi:10.1117/12.789285
\bibitem[Lehmann~{\it et al.}(2005)]{2005AA...431..847L} Lehmann I., Becker T., Fabrika S., Roth M., Miyaji T., Afanasiev V., Sholukhova O., et al., 2005, A\&A, 431, 847. doi:10.1051/0004-6361:20035827
\bibitem[L{\'o}pez-Hern{\'a}ndez~{\it et al.}(2013)]{2013MNRAS.430..472L} L{\'o}pez-Hern{\'a}ndez J., Terlevich E., Terlevich R., Rosa-Gonz{\'a}lez D., D{\'\i}az {\'A}., Garc{\'\i}a-Benito R., V{\'\i}lchez J., et al., 2013, MNRAS, 430, 472. doi:10.1093/mnras/sts658
\bibitem[L{\'o}pez~{\it et al.}(2008)]{2008MNRAS.384..464L} L{\'o}pez R., S{\'a}nchez S.~F., Garc{\'\i}a-Lorenzo B., G{\'o}mez G., Estalella R., Riera A., Busquet G., 2008, MNRAS, 384, 464. doi:10.1111/j.1365-2966.2007.12718.x
\bibitem[Monreal-Ibero~{\it et al.}(2005)]{2005ApJ...628L.139M} Monreal-Ibero A., Roth M.~M., Sch{\"o}nberner D., Steffen M., B{\"o}hm P., 2005, ApJL, 628, L139. doi:10.1086/432664
\bibitem[Montero~{\it et al.}(2023)]{2023IAUS384Montero} Montero H. et al., 2023, these proceedings
\bibitem[Morrissey~{\it et al.}(2018)]{2018ApJ...864...93M} Morrissey P., Matuszewski M., Martin D.~C., et al., 2018, ApJ, 864, 93. doi:10.3847/1538-4357/aad597
\bibitem[Roth~{\it et al.}(1998)]{1998SPIE.3355..798R} Roth M.~M., Bauer S.-M., Dionies F., Fechner T., et al., 1998, SPIE, 3355, 798. doi:10.1117/12.316793
\bibitem[Roth \& Laux(1998)]{1998ASPC..152..168R} Roth M.~M., Laux U., 1998, ASPC, 152, 168
\bibitem[Roth~{\it et al.}(2004a)]{2004AN....325..147R} Roth M.~M., Becker T., B{\"o}hm P., Kelz A., 2004, AN, 325, 147. doi:10.1002/asna.200310196
\bibitem[Roth~{\it et al.}(2004b)]{2004ApJ...603..531R} Roth M.~M., Becker T., Kelz A., Schmoll J., 2004, ApJ, 603, 531. doi:10.1086/381526
\bibitem[Roth~{\it et al.}(2005)] {2005PASP..117..620R}Roth M.~M., Kelz A., Fechner T., Hahn T., Bauer S.-M., Becker T., B{\"o}hm P., et al., 2005, PASP, 117, 620. doi:10.1086/429877
\bibitem[Roth~{\it et al.}(2006)]{2006pnbm.conf..239R} Roth M.~M., Becker T., B{\"o}hm P., Sch{\"o}nberner D., Steffen M., Exter K., 2006, pnbm.conf, 239. doi:10.1007/3-540-34270-2\_36
\bibitem[Roth~{\it et al.}(2010)]{2010SPIE.7742E..09R} Roth M.~M., Fechner T., Wolter D., Sandin C., Kelz A., Bauer S.~M., Popow E., et al., 2010, SPIE, 7742, 774209. doi:10.1117/12.856936
\bibitem[Roth~{\it et al.}(2018)]{2018A&A...618A...3R} Roth M.~M., Sandin C., Kamann S., Husser T.-O., Weilbacher P.~M., Monreal-Ibero A., Bacon R., et al., 2018, A\&A, 618, A3. doi:10.1051/0004-6361/201833007
\bibitem[Roth~{\it et al.}(2019)]{2019AN....340..989R} Roth M.~M., Weilbacher P.~M., Castro N., 2019, AN, 340, 989. doi:10.1002/asna.201913751
\bibitem[Roth~{\it et al.}(2021)]{2021ApJ...916...21R} Roth M.~M., Jacoby G.~H., Ciardullo R., Davis B.~D., Chase O., Weilbacher P.~M., 2021, ApJ, 916, 21
\bibitem[S{\'a}nchez~{\it et al.}(2012)]{2012AA...538A...8S}S{\'a}nchez S.~F., Kennicutt R.~C., Gil de Paz A., van de Ven G., V{\'\i}lchez J.~M., Wisotzki L., Walcher C.~J., et al., 2012, A\&A, 538, A8. doi:10.1051/0004-6361/201117353
\bibitem[Sandin~{\it et al.}(2008)]{2008AA...486..545S} Sandin C., Sch{\"o}nberner D., Roth M.~M., Steffen M., B{\"o}hm P., Monreal-Ibero A., 2008, A\&A, 486, 545. doi:10.1051/0004-6361:200809635
\bibitem[Sandin~{\it et al.}(2010)]{2010A&A...512A..18S} Sandin C., Jacob R., Sch{\"o}nberner D., Steffen M., Roth M.~M., 2010, A\&A, 512, A18. doi:10.1051/0004-6361/200911796
\bibitem[Sil'chenko \& Afanasiev(2000)]{2000AJ....119..126A} Afanasiev V.~L., Sil'chenko O.~K., 2000, AJ, 119, 126. doi:10.1086/301162
\bibitem[Soemitro~{\it et al.}(2023a)] {2023A&A...671A.142S} Soemitro A.~A., Roth M.~M., Weilbacher P.~M., Ciardullo R., Jacoby G.~H., et al., 2023a, A\&A, 671, A142
\bibitem[Soemitro~{\it et al.}(2023b)]{2023IAUS384Soemitro} Soemitro A. et al., 2023b, these proceedings
\bibitem[Steidel~{\it et al.}(2022)]{2022SPIE12184E..23S} Steidel C., Peng E., Fucik J., et al., 2022, SPIE, 12184, 1218423. doi:10.1117/12.2629464
\bibitem[Stuik~{\it et al.}(2006)]{2006NewAR..49..618S} Stuik R., Bacon R., Conzelmann R., Delabre B., Fedrigo E., Hubin N., Le Louarn M., et al., 2006, NewAR, 49, 618. doi:10.1016/j.newar.2005.10.015
\bibitem[Stetson~(1987)]{1987PASP...99..191S} Stetson P.~B., 1987, PASP, 99, 191. doi:10.1086/131977
\bibitem[Tyson~(1990)]{1990ASPC...10..292T} Tyson J.~A., 1990, ASPC, 10, 292
\bibitem[Vanderriest~{\it et al.}(1987)]{1987AnPh...12..207V} Vanderriest C., Haddad B., Lemonnier J.~P., 1987, AnPh, 12, 207
\bibitem[Weilbacher~{\it et al.}(2020)]{2020A&A...641A..28W} Weilbacher P.~M., Palsa R., Streicher O., et al., 2020, A\&A, 641, A28. doi:10.1051/0004-6361/202037855
\bibitem[Weitzel~{\it et al.}(1996)]{1996AAS..119..531W} Weitzel L., Krabbe A., Kroker H., et al., 1996, A\&AS, 119, 531
\bibitem[Wisotzki~{\it et al.}(2003)]{2003AA...408..455W} Wisotzki L., Becker T., Christensen L., Helms A., Jahnke K., Kelz A., Roth M.~M., et al., 2003, A\&A, 408, 455. doi:10.1051/0004-6361:20031004
\end{thebibliography}
\end{document}